\documentclass[a4paper,14pt]{article}
\usepackage[english]{babel}
\pdfoutput=1 

\usepackage{jheppub} 

\usepackage[T1]{fontenc} 

\usepackage[T1,T2A]{fontenc}
\usepackage[utf8x]{inputenc}

\usepackage{comment}

\usepackage{cases}

\newcommand{\bi}{\begin{itemize}}
    \newcommand{\ei}{\end{itemize}}
\newcommand{\bea}{\begin{eqnarray}}
    \newcommand{\eea}{\end{eqnarray}}
\newcommand{\bt}{\begin{tabular}}
    \newcommand{\et}{\end{tabular}}
\newcommand{\bc}{\begin{center}}
    \newcommand{\ec}{\end{center}}

\newcommand{\be}{\begin{equation}}
    \newcommand{\ee}{\end{equation}}
\newcommand{\ba}{\begin{array}}
    \newcommand{\ea}{\end{array}}

\def\bbox{{\,\lower0.9pt\vbox{\hrule \hbox{\vrule height 0.2 cm
                \hskip 0.2 cm \vrule height 0.2 cm}\hrule}\,}}
\newcommand{\dsl}{\pa \kern-0.5em /}

\makeatletter \@addtoreset{equation}{section} \makeatother

\def\slashchar#1{\setbox0=\hbox{$#1$}           
    \dimen0=\wd0                                 
    \setbox1=\hbox{/} \dimen1=\wd1               
    \ifdim\dimen0>\dimen1                        
    \rlap{\hbox to \dimen0{\hfil/\hfil}}      
    #1                                        
    \else                                        
    \rlap{\hbox to \dimen1{\hfil$#1$\hfil}}   
    /                                         
    \fi}

\pdfoutput=1

\title{\boldmath  $\mathcal{N}=2$ superconformal gravitino in harmonic superspace}

\author[a,b]{Evgeny~Ivanov,}
\author[a,b]{Nikita~Zaigraev}

\affiliation[a]{Bogoliubov Laboratory of Theoretical Physics, JINR,\\141980 Dubna, Moscow region, Russia}
\affiliation[b]{Moscow Institute of Physics and Technology,\\ 141700 Dolgoprudny, Moscow region, Russia}

\emailAdd{eivanov@theor.jinr.ru}
\emailAdd{nikita.zaigraev@phystech.edu}

\abstract{We present the harmonic superspace formulation of $\mathcal{N}=2$ gravitino multiplet,
the simplest $\mathcal{N}=2$ half-integer spin gauge supermultiplet. It is shown that, quite similar to other $\mathcal{N}=2$ gauge multiplets,
the gravitino supermultiplet is described by unconstrained analytic prepotentials $h^{++\alpha}$ and $h^{+++}$ which contain a conformal gravitino in the Wess-Zumino gauge.
The analytic prepotentials naturally come out from the study of $\mathcal{N}=2$
supercurrents associated with hidden symmetries of $\mathcal{N}=2$ vector-hypermultiplet system. We construct the
covariant $\mathcal{N}=2$ superfield strengths and the invariant $\mathcal{N}=2$ superfield actions and sketch
their component contents. We observe that, at cost of introducing new auxiliary
coordinates $\Psi^\alpha$ and $\omega^+$, the gravitino analytic prepotentials acquire a nice geometric interpretation as
extra vielbeins of the covariant harmonic derivative $\mathfrak{D}^{++}$. We speculate on a possible origin of the
additional coordinates, including  their relationship with $\mathcal{N}=2$ supertwistors. }

\arxivnumber{2412.14822}

\makeatletter
\gdef\@fpheader{}
\makeatother

\begin{document}

\maketitle
\flushbottom


\section{Introduction}\label{sec: intro}

The natural geometric description of all ${\cal N}=2, 4D$ theories of interest is achieved within the harmonic superspace (HSS) approach.
The harmonic superspace \cite{Galperin:1984av, 18}, besides the ordinary $\mathcal{N}=2$ superspace coordinates
$z = \{x^{\alpha\dot{\alpha}}, \theta^{\alpha i}, \bar{\theta}^{\dot{\alpha} i} \}$, involves auxiliary harmonic coordinates $u^{\pm}_i$, $u^{+i} u^-_i =1$.
The adding of harmonic coordinates gives rise to the two radical distinctions from the ordinary $\mathcal{N}=2$ superspace description. First of all, it is the presence
of a new supersymmetric invariant superspace -- \textit{analytic superspace},
with coordinates $\zeta = \{x^{\alpha\dot{\alpha}}_A, \theta^{+\alpha}, \bar{\theta}^{+\dot{\alpha}}, u^\pm_i\}$, where analytic coordinates are defined as
\begin{equation}
x_A^{\alpha\dot{\alpha}} = x^{\alpha\dot{\alpha}}- 4i \theta^{\alpha (i} \bar{\theta}^{\dot{\alpha}j)} u^+_i u^-_j,
\quad
 \theta^{+\alpha} =\theta^{\alpha i} u^+_i ,
 \quad
  \bar{\theta}^{+\dot{\alpha}} = \bar{\theta}^{\dot{\alpha} i} u^+_i.
\end{equation}
The second fundamental difference is the presence of \textit{harmonic derivatives} $\partial^{\pm\pm} = u^{\pm i}\frac{\partial}{\partial u^{\mp i}}$, which in the analytic basis take the form:
\begin{equation}
    \mathcal{D}^{\pm\pm} = \partial^{\pm\pm} - 4i \theta^{\alpha\pm} \bar{\theta}^{\dot{\alpha}\pm} \partial_{\alpha\dot{\alpha}} + \theta^{\pm\hat{\alpha}} \partial^{\pm}_{\hat{\alpha}},
    \quad \hat{\alpha} = (\alpha, \dot\alpha).
\end{equation}
All known gauge $\mathcal{N}=2$ supermultiplets are described by the appropriate analytic prepotentials, with gauge transformations involving harmonic derivatives. The off-shell structure of the dynamical actions is
determined through solving the so called
 harmonic zero-curvature equations.

The analytic gauge  prepotentials have a natural geometric origin in the harmonic approach.  For example, the prepotentials underlying $\mathcal{N}=2$ Einstein supergravity
 are vielbeins that covariantize the harmonic derivative with respect to the analyticity-preserving superdiffeomorphisms \cite{Galperin:1987em}:
\begin{equation}\label{eq: Ein}
    \mathcal{D}^{++}
    \quad
    \to
    \quad
    \mathfrak{D}^{++}_{Ein}
    =
    \mathcal{D}^{++} + h^{++\alpha\dot{\alpha}}\partial_{\alpha\dot{\alpha}}
    +
    h^{++\hat{\alpha}+} \partial^-_{\hat{\alpha}}
    +
    h^{++5} \partial_5.
\end{equation}
Analogously, the analytic prepotentials of $\mathcal{N}=2$ conformal supergravity appear in  $\mathfrak{D}^{++}$ as \cite{Galperin:1987ek}:
\begin{equation}\label{eq: Conf}
    \mathcal{D}^{++}
    \quad
    \to
    \quad
    \mathfrak{D}^{++}_{conf}
    =
    \mathcal{D}^{++} + h^{++\alpha\dot{\alpha}}\partial_{\alpha\dot{\alpha}}
        +
    h^{++\hat{\alpha}+} \partial^-_{\hat{\alpha}}
    +
    h^{(+4)} \partial^{--}.
\end{equation}
At the component level, the set of analytic prepotentials $h^{++M}(\zeta)$  contains the fields of off-shell $\mathcal{N}=2$ supergravity Weyl  multiplet,
together with an infinite number of pure gauge degrees of freedom. One of the basic advantages of the harmonic approach to $\mathcal{N}=2$ supergravity
is that this geometric formulation admits a natural generalization to $\mathcal{N}=2$ higher spins.
The higher-spin  analytic prepotentials have a similar geometric interpretation for any {\it integer} $\mathcal{N}=2$ spin $\mathbf{s}$\footnote{We use bold $\mathbf{s}$
to denote the $\mathcal{N}=2$ supermultiplet with highest physical spin $s$.}  in both Einstein \cite{Buchbinder:2021ite, Buchbinder:2022kzl, Buchbinder:2022vra}
and conformal cases \cite{Buchbinder:2024pjm}\footnote{Superconformal higher-spin interactions of hypermultiplet were also considered in the framework of $\mathcal{N}=2$ projective superspace \cite{Kuzenko:2024vms}.}.
Moreover, such a geometric covariantization of the harmonic derivatives immediately fixes the cubic interaction
$(\mathbf{s}, \mathbf{\frac{1}{2}}, \mathbf{\frac{1}{2}})$ of the $\mathcal{N}=2$ spin $\mathbf{s}$ gauge multiplets with the fundamental $\mathcal{N}=2$ matter multiplet, the hypermultiplet.

Until now, the harmonic  superspace formulation of the {\it half-integer} $\mathcal{N}=2$ higher spins is not known.
Our paper can be regarded as a step towards filling this gap.
We study the simplest example of such a multiplet --  the  $\mathcal{N}=2$ gravitino multiplet\footnote{A superfield description of $\mathcal{N}=1$ gravitino supermultiplet was constructed by V. Ogievetsky and E. Sokatchev in the seminal papers
    \cite{Ogievetsky:1975vk, Ogievetsky:1976qb} (see also \cite{Buchbinder:1998qv} for a pedagogical review). The superfield approach to
    $\mathcal{N}=1$ higher-spin supermultiplets with an arbitrary integer superspin was elaborated by  S. Kuzenko and A. Sibiryakov \cite{Kuzenko:1993jq}.},
i.e. the spin $\mathbf{\frac{3}{2}}$ multiplet.

Let us firstly argue why the structure of such a supermultiplet should be different from that of the integer $\mathcal{N}=2$ higher spin ones and
cannot be obtained through  a direct covariantization of the harmonic derivative $\mathcal{D}^{++}$. To start with, $\mathcal{N}=2$ gravitino supermultiplet
should contain a spin $\frac{3}{2}$ field, which undergoes a gauge transformation with spinor parameter.  An important point is that the latter should be a singlet with respect
to the standard $\mathcal{N}=2$ $R$ symmetry $SU(2)$.
Indeed, the hypermultiplet reveals an invariance under the standard
rigid $\mathcal{N}=2$ supersymmetry, and the gauging of the latter naturally leads to an interaction involving  $\mathcal{N}=2$ supergravity multiplet
 \cite{Buchbinder:2022kzl, Buchbinder:2022vra, Buchbinder:2024pjm} through lengthening the harmonic derivative, like in \eqref{eq: Ein} and \eqref{eq: Conf}.
So, in order to generate coupling with $\mathcal{N}=2$ gravitino multiplet, which is different from  the $\mathcal{N}=2$ supergravity one, we need another type of rigid supersymmetry transformations.
Since the hypermultiplet action on its own does not reveal invariances with respect to additional supersymmetries, we conclude that the $\mathcal{N}=2$ gravitino multiplet cannot
be recovered by straightforwardly covariantizing the harmonic derivative in the hypermultiplet action (and so there exists no minimal cubic $(\mathbf{\frac{3}{2}}, \mathbf{\frac{1}{2}}, \mathbf{\frac{1}{2}})$
coupling, in agreement with the Metsaev classification \cite{Metsaev:2019aig}). As a result, we need to consider $\mathcal{N}=2$ actions with some extra supersymmetry.
The simplest example of this kind is supplied by the ``hidden'' supersymmetry of  $\mathcal{N}=2$ vector-hypermultiplet complex.

In this article, we construct the harmonic superspace formulation of $\mathcal{N}=2$ conformal gravitino supermultiplet, based on the reasoning just adduced.
We  start with hidden rigid supersymmetry of $\mathcal{N}=2$ vector-hypermultiplet system and derive the corresponding conserved $\mathcal{N}=2$ supercurrents.
Then, following the approach of Ogievetsky and Sokatchev to  ${\cal N}=1$ supergravity \cite{Ogievetsky:1976qc},
we identify analytic $\mathcal{N}=2$ gravitino prepotentials as gauge superfields which couple to these $\mathcal{N}=2$ supercurrents (at the linearized level). Finally, we suggest  a geometric interpretation
of these prepotentials as vielbeins associated with some extra coordinates extending the standard analytic ${\cal N}=2$ superspace.

\medskip

Another approach to a superfield description of $\mathcal{N}=2$ superconformal
gravitino multiplet was developed in ref. \cite{Hutchings:2023iza}, based on  the $\mathcal{N}=2$ reduction
of $\mathcal{N}=3$ supercurrent. It can be contrasted with ours which proceeds from symmetries
of $\mathcal{N}=2$ off-shell harmonic superspace actions. It naturally leads to an unconstrained off-shell
cubic $(\mathbf{\frac{3}{2}}, \mathbf{\frac{1}{2}}, \mathbf{\frac{1}{2}})$  vertex which is a necessary ingredient for further construction
of $\mathcal{N}=2$  non-conformal  gravitino theory. In section \ref{eq: sec 6} we demonstrate, that the superconformal gravitino
prepotential introduced in \cite{Hutchings:2023iza} can be recovered as a harmonic-independent gauge-fixed form
of the Mezincescu-type prepotential in our system.

\section{Hidden supersymmetries, $R$-symmetries and $\mathcal{N}=2$ supercurrents}

In this section we discuss rigid symmetries of free massless $\mathcal{N}=4$ vector multiplet action in the ${\cal N}=2$ superfield formulation \cite{Buchbinder:2020yvf}
and their gauging. In $\mathcal{N}=2$ harmonic superspace this action  amounts to a sum of actions for free massless $\mathcal{N}=2$ vector multiplet and hypermultiplet:
\begin{equation}\label{eq: action}
    S^{\mathcal{N}=4} = \frac{1}{4} \int d^4x d^8 \theta du\;  V^{++} V^{--}
    -
    \frac{1}{2} \int d\zeta^{(-4)} q^{+a} \mathcal{D}^{++} q^+_a.
\end{equation}
Here $q^{+a} = (\tilde{q}^+, q^+), \,\widetilde{q^{+a}} = - \epsilon_{ab}q^{+b}$ and $V^{--}$ is defined as the solution of zero curvature equation $\mathcal{D}^{++} V^{--} = \mathcal{D}^{--} V^{++}$.
The $\mathcal{N}=2$ vector multiplet action reveals the gauge freedom $\delta_\lambda^{(0)} V^{++} = \mathcal{D}^{++} \lambda$
with an unconstrained analytic parameter $\lambda(\zeta)$. The action \eqref{eq: action} possesses an extra invariance under the additional internal group
$SU(2)_{PG}$ (``Pauli-G\"ursey'') acting on the doublet indices $a$ of the hypermultiplet.

\medskip

The total action \eqref{eq: action} is also invariant under global transformations of hidden ${\cal N}=2$ supersymmetry\footnote{Here $a= (1,2)$
and $\varepsilon^{\alpha a} = \epsilon^{ab} \varepsilon^{\alpha}_b$, $\varepsilon^\alpha_a = \epsilon_{ab} \varepsilon^{\alpha b}$. We denote the parameters of hidden supersymmetry
as $\varepsilon^{\alpha a}$, and the parameters of the explicitly realized ${\cal N}=2$ supersymmetry as $\epsilon^{\alpha i}$.}
\begin{equation}\label{eq: hidden susy}
    \begin{split}
    &\delta_\varepsilon V^{++} = ( \varepsilon^{\alpha a} \theta^+_\alpha + \bar{\varepsilon}^a_{\dot{\alpha}} \bar{\theta}^{+\dot{\alpha}}) q^+_a,
    \\
    &\delta_\varepsilon q^+_a = - \frac{1}{2} (\mathcal{D}^+)^4 \left[ ( \varepsilon^{\alpha}_{a} \theta^-_\alpha + \bar{\varepsilon}_{a\dot{\alpha}} \bar{\theta}^{-\dot{\alpha}})  V^{--} \right],
    \end{split}
\end{equation}
with rigid fermionic parameters $\varepsilon^{\alpha a}$ and $\bar{\varepsilon}^{\dot{\alpha}}_{a} = \widetilde{\varepsilon^{\alpha a}}$.

The analytic gauging of the transformation parameter, $\epsilon^{\alpha a} \to \lambda^{\alpha a}(\zeta)$, leads
to the following variation of the action\footnote{We define the integration measure as $\int d^4x d^8 \theta du = \int d\zeta^{(-4)} (\mathcal{D}^+)^4$,
    $\int d\zeta^{(-4)} = \int d^4x (\mathcal{D}^-)^4 du$, and use the standard notation $
    (\mathcal{D}^\pm)^4 = \frac{1}{16} (\mathcal{D}^\pm)^2 (\bar{\mathcal{D}}^\pm)^2
    $ .}:
\begin{equation}
    \delta_\lambda S^{\mathcal{N}=4} =  \frac{1}{2} \int d\zeta^{(-4)}\; \mathcal{D}^{++} \lambda^\alpha_a  \; (\mathcal{D}^+)^4 \left( q^{+a} \theta^-_\alpha V^{--} \right).
\end{equation}
Using this variation, we define the conserved analytic supercurrent\footnote{We use  $\approx$ to denote the equality modulo the equations of motion.},
\begin{equation} \label{DefJalpha}
    \mathcal{J}_\alpha^{++a}
    =
    \frac{1}{2}
    (\mathcal{D}^+)^4 \left( q^{+a} \theta^-_\alpha V^{--} \right)
    =
    \frac{1}{32} q^{+a} (\mathcal{D}^+)^2 \left( \theta^-_\alpha \mathcal{W} \right)
    \approx
        \frac{1}{16} q^{+a} \mathcal{D}^+_\alpha \mathcal{W},
\end{equation}
which satisfies the on-shell ``conservation'' condition:
\begin{equation}\label{eq: cons 1}
    \mathcal{D}^{++}    \mathcal{J}_\alpha^{++a} \approx 0
\end{equation}
(in \eqref{DefJalpha} we have defined the chiral ${\cal N}=2$ gauge field strength $\mathcal{W} \sim (\bar{D}^+)^2 V^{--}$).
This supercurrent corresponds to hidden supersymmetry of the action \eqref{eq: action},
so it can naturally couple to the analytic gauge prepotential, which contains a spin $\frac{3}{2}$ field.
However, before gauging the hidden supersymmetry, it is useful to consider some other symmetries of
the action \eqref{eq: action}.

\medskip

This  another class of global symmetries of \eqref{eq: action} is formed by the bosonic transformations:
\begin{equation}\label{eq: hiden R}
    \begin{split}
        &\delta_r V^{++} =  2 \left(r^{-a} (\theta^+)^2 + \bar{r}^{-a} (\bar{\theta}^+)^2 \right) q^+_a,
        \\
        &\delta_r q^+_a =   -\frac{1}{2} (\mathcal{D}^+)^4 \Big[ \Big\{ r^+_a (\theta^-)^2 - 2 r^-_a (\theta^+ \theta^-) + \bar{r}^+_a (\bar{\theta}^-)^2 -2 \bar{r}^-_a (\bar{\theta}^+ \bar{\theta}^-) \Big\} V^{--} \Big].
    \end{split}
\end{equation}
Here rigid parameters are defined as $r^{\pm a} = r^{a i } u^\pm_i$.  These transformations generate the hidden $R$-symmetries of the action \eqref{eq: action} which
extend the explicit $R$ symmetry group of \eqref{eq: action} from $U(2)_R\times SU(2)_{PG}$
to $SU(4)$ (which is just the $R$-symmetry group $SU(4)_R$ of the complete ${\cal N}=4$ supersymmetry). The on-shell closure of these hidden $R$-symmetry transformations yields the linear
$U(2)_R$ and $SU(2)_{PG}$ transformations, in agreement with the property that they form the coset
$SU(4)_R/[SU(2)_R \times U(1)_R \times SU(2)_{PG}]$ part of the full automorphism symmetry $SU(4)_R$ of $\mathcal{N} = 4$
supersymmetry algebra \cite{Buchbinder:2020yvf}.

There are two possible types of gauging of the transformations  \eqref{eq: hiden R}:\\

\textbf{1.} The analytic localization $r^+_a \to \lambda^{+a}(\zeta)$, $r^-_a \to 0$ yields:
\begin{equation}
    \begin{split}
        \delta_{\lambda+} S^{\mathcal{N}=4} =&
      \frac{1}{2} \int d^4\zeta^{(-4)} \Big\{ (\mathcal{D}^{++}\lambda^+_a) \underbrace{q^{+a} (\mathcal{D}^+)^4 \left[ (\theta^-)^2 V^{--}\right]}_{2\mathcal{J}^{+a}}
        + 2 \lambda^+_a
        \underbrace{q^{+a} (\mathcal{D}^+)^4 \left[ (\theta^+ \theta^-) V^{--} \right]}_{2 \theta^{+\alpha} \mathcal{J}^{++a}_\alpha }
        \Big\}.
    \end{split}
\end{equation}
It ensures the existence of the analytic $\mathcal{N}=2$ supercurrent,
\begin{equation}
    \mathcal{J}^{+a} = \frac{1}{2}(\mathcal{D}^+)^4 \left( q^{+a} (\theta^-)^2 V^{--}  \right)
    =
    \frac{1}{32} (\mathcal{D}^+)^2 \left( q^{+a} (\theta^-)^2 \mathcal{W} \right)
    \approx
    \frac{1}{8} q^{+a} \mathcal{W}
    +
    \frac{1}{8} q^{+a} \theta^{-\alpha} \mathcal{D}^+_\alpha \mathcal{W},
\end{equation}
with the conservation conditions:
\begin{equation}\label{eq: R current cons}
    \mathcal{D}^{++}    \mathcal{J}^{+a}  \approx 2 \theta^{+\alpha} \mathcal{J}_\alpha^{++a},
    \qquad
    (\mathcal{D}^{++})^2    \mathcal{J}^{+a}  \approx 0.
\end{equation}
Since this supercurrent corresponds to the hidden $R$-symmetries, we expect that the associate gauge superfield prepotential contains the relevant $R$-symmetry gauge fields.\\

\textbf{2.} The analytic localization $r^{-a} \to \lambda^{-a}(\zeta)$,  $r^{+a} \to 0$ leads to the variation:
\begin{equation}
    \delta_{\lambda^-} S^{\mathcal{N}=4} =  - \int d\zeta^{(-4)} \; \mathcal{D}^{++} \lambda^-_a \underbrace{q^{+a} (\mathcal{D}^+)^4 \left[  (\theta^+\theta^-) V^{--}  \right]}_{2\theta^{+\alpha} \mathcal{J}^{++a}_\alpha}.
\end{equation}
The $\mathcal{N}=2$ supercurrent obtained in this way is given by $\theta^{+\alpha} \mathcal{J}^{++a}_\alpha$, and it is the on-shell descendant of the supercurrent $\mathcal{J}^{+a}$ (see \eqref{eq: R current cons}).
This implies that the transformations $\delta_{\lambda^+}$ in fact amount to the most general analytic gauging of   \eqref{eq: hiden R}.

\medskip

If we restrict the transformation parameters as $\lambda^{+}_{a} = \mathcal{D}^{++} \lambda^{-}_{a}$ and $\mathcal{D}^{++}\lambda^+_a =0$,
the total transformation $\delta_{\lambda^+} + \delta_{\lambda^-}$ yields just the hidden  rigid $R$-symmetry \eqref{eq: hiden R}.

\medskip

The analytic supercurrents $\mathcal{J}^+_a$ and $\mathcal{J}^{++}_{a \alpha}$ are related by the rigid $\mathcal{N}=2$ supersymmetry  transformations realized in the harmonic superspace as
\begin{equation}\label{eq: R-sym}
    \delta_\epsilon \mathcal{J}^{+}_a = 2 \epsilon^{-\alpha}  \mathcal{J}^{++}_{\alpha a},
    \qquad
    \delta_\epsilon  \mathcal{J}^{++}_{\alpha a} = 0,
\end{equation}
and by the conservation conditions \eqref{eq: R current cons}.

\medskip

In addition to the supercurrents just defined, it is useful to consider the ``master''  (conformal) $\mathcal{N}=2$ supercurrent:
\begin{equation}\label{eq: principle sc}
    \mathfrak{J}^{+a} = \frac{1}{8} q^{+a} \mathcal{W}.
\end{equation}
This supercurrent corresponds to the supersymmetry invariant part of $\mathcal{J}^{+a}$ and is subject to the conservation conditions:
\begin{equation}\label{eq: psc cons equation}
    \begin{cases}
        \mathcal{D}^{++}    \mathfrak{J}^{+a} \approx 0,
        \\
        (\mathcal{D}^+)^2   \mathfrak{J}^{+a} \approx 0,
        \\
        \bar{\mathcal{D}}^+_{\dot{\alpha}} \mathfrak{J}^{+a} = 0.
    \end{cases}
\end{equation}
In contrast to the analytic supercurrents $\mathcal{J}^{++}_{\alpha a}$ and $\mathcal{J}^+_a$,  the master supercurrent  $\mathfrak{J}^{+}_a$
is not analytic, $\mathcal{D}^+_\alpha\mathfrak{J}^{+}_a \neq 0$, but it still
obeys the ``half-analyticity'' condition $\bar{\mathcal{D}}^+_{\dot{\alpha}}\mathfrak{J}^{+}_a = 0$.
Another useful property is that the analytic  supercurrents $\mathcal{J}^{+a}$ and $\mathcal{J}^{++a}_\alpha$ can be constructed as descendants of $  \mathfrak{J}^{+a}$:
\begin{equation}\label{eq: ac pc}
    \mathcal{J}^{++a}_\alpha
    =
    \frac{1}{2} \mathcal{D}^+_\alpha    \mathfrak{J}^{+a},
    \qquad
    \mathcal{J}^{+a} =  \mathfrak{J}^{+a}  -  \theta^{-\alpha}  \mathcal{D}^+_\alpha \mathfrak{J}^{+a}.
\end{equation}
Using the hidden supersymmetry transformations \eqref{eq: hidden susy}, one can restore  the $\mathcal{N}=4$ supercurrent multiplet in terms of $\mathcal{N}=2$ harmonic superfields. For example, the variation
\begin{equation}\label{eq: hidden susy J}
    \delta_{\bar{\varepsilon}} \mathfrak{J}^{+a} = \frac{1}{16} \bar{\varepsilon}^{\dot{\alpha}a} \bar{\mathcal{D}}^+_{\dot{\alpha}} \left( q^{+a} \mathcal{D}^{--}  q^+_a+ \frac{1}{2} \mathcal{W} \bar{\mathcal{W}} \right)
    =
    \frac{1}{16} \bar{\varepsilon}^{\dot{\alpha}a} \bar{\mathcal{D}}^+_{\dot{\alpha}} \mathcal{J},
\end{equation}
gives rise to $\mathcal{N}=2$ supercurrent $\mathcal{J}$, which couples to the prepotentials of $\mathcal{N}=2$ conformal supergravity \cite{Zaigraev:2024xve, Butter:2010sc, Kuzenko:1999pi}.
Using the relations \eqref{eq: ac pc}, one can readily  determine the explicit form of the transformations of the analytic descendants.

\section{Gauge prepotentials and cubic $(\mathbf{\frac{3}{2}}, \mathbf{1}, \mathbf{\frac{1}{2}})$ coupling}

Based on the conservation laws \eqref{eq: cons 1} and \eqref{eq: R current cons}, it is natural to introduce the unconstrained analytic prepotentials $h^{++\alpha}_a$ and $h^{+++}_a$ with the gauge transformations:
\begin{equation}\label{eq: GFa}
    \delta_\lambda h_a^{++\alpha} = \mathcal{D}^{++} \lambda_a^\alpha
    +
    2 \theta^{+\alpha} \lambda^+_a,
    \qquad
    \delta_\lambda h^{+++}_a = \mathcal{D}^{++}\lambda^+_a,
\end{equation}
as well as their tilde-conjugates. The parameters $\lambda_a^{\alpha}(\zeta)$ and $\lambda_a^+(\zeta)$ are unconstrained analytic superfields.

Now one can construct the cubic vertex
\begin{equation}\label{eq: cubic vertex}
    S_{int} = - \kappa \int d\zeta^{-4} \; \Big[ h^{+++a} \mathcal{J}^+_a
    +
    h^{++\alpha a} \mathcal{J}^{++}_{\alpha a}
    +
    (\;\;  \widetilde{}\;\; \text{conj})
    \Big].
\end{equation}
The requirement of invariance under rigid $\mathcal{\mathcal{N}}=2$ supersymmetry leads to the following transformations
of the analytic prepotentials\footnote{Under the transformations of hidden ${\cal N}=2$ supersymmetry $\delta_{\varepsilon}$, these prepotentials transform thorough
the prepotentials of $\mathcal{N}=2$ supergravity, as can be seen, e.g.,  from the form of the variation \eqref{eq: hidden susy J}.
This is in agreement with the structure of $\mathcal{N}=4$ supergravity multiplet.}:
\begin{equation}\label{eq: susy transf}
    \delta_\epsilon \mathcal{J}^+_a =  2 \epsilon^{-\alpha} \mathcal{J}^{++}_{\alpha a},
    \;\;\;
    \delta_\epsilon \mathcal{J}^{++}_{\alpha a} = 0
    \quad
    \Rightarrow
    \quad
    \delta_\epsilon h^{++\alpha a} = - 2 \epsilon^{-\alpha}h^{+++a},
    \;\;\;
    \delta_\epsilon h^{+++ a} = 0.
\end{equation}

One can consider the following gauge transformations of the matter superfields $V^{++}, q^{+a}$ describing the $\mathcal{N}=2$ spin $\mathbf{1}$ and spin $\mathbf{\frac{1}{2}}$ supermultiplets:
\begin{subequations}\label{eq: matter GF}
\begin{equation}
    \delta^{(1)}_\lambda V^{++} = \kappa ( \lambda^{\alpha a} \theta^+_\alpha + \bar{\lambda}^a_{\dot{\alpha}} \bar{\theta}^{+\dot{\alpha}}) q^+_a,
    \qquad
    \delta^{(1)}_\lambda q^+_a = - \frac{1}{2} (\mathcal{D}^+)^4 \left[ ( \lambda^{\alpha}_{a} \theta^-_\alpha + \bar{\lambda}_{a\dot{\alpha}} \bar{\theta}^{-\dot{\alpha}})  V^{--} \right],
\end{equation}
\begin{equation}
    \delta^{(2)}_\lambda V^{++}  = 0,
    \qquad\qquad\qquad
    \delta^{(2)}_\lambda q^+_a = - \frac{\kappa}{2} (\mathcal{D}^+)^4 \left[ \lambda^+_a (\theta^-)^2 V^{--} + \bar{\lambda}^+_a (\bar{\theta}^-)^2 V^{--} \right].
\end{equation}
\end{subequations}
The first type of transformations, $\delta_\lambda^{(1)}$, is the  analytic localization of hidden supersymmetry  \eqref {eq: hidden susy}.
The second type transformations, $\delta_\lambda^{(2)}$, corresponds to the analytic localization  of $R$-symmetry transformations \eqref{eq: R-sym}.
The total action $S^{\mathcal{N}=4} + S_{int}$ should be gauge invariant in the first order in the coupling constant $\kappa$ under the gauge transformations  \eqref{eq: matter GF}
and the appropriate compensating gauge transformations of the analytic prepotentials. These latter transformations are just given by \eqref{eq: GFa}.

Due to the gauge invariance (in zero order in $\kappa$) of the supercurrents,
\begin{equation}
    \delta_\lambda^{(0)} \mathcal{J}^{++a}_\alpha =     \frac{1}{2}
    (\mathcal{D}^+)^4 \left( q^{+a} \theta^-_\alpha \mathcal{D}^{--} \lambda \right)
    = 0,
    \qquad
    \delta_\lambda^{(0)} \mathcal{J}^+_a
    =
    \frac{1}{2}(\mathcal{D}^+)^4 \left( q^{+a} (\theta^-)^2 \mathcal{D}^{--} \lambda  \right)
= 0\,,
\end{equation}
the   cubic $(\mathbf{\frac{3}{2}}, \mathbf{1}, \mathbf{\frac{1}{2}})$ coupling \eqref{eq: cubic vertex} is consistent to the leading order in $\kappa$.

\section{Wess-Zumino type gauge}

To reveal the physical contents of the gauge prepotentials  defined  above it is useful to fix the $SU(2)_{PG}$ doublet index as $a=1$ and to study the
prepotentials $h^{++\alpha} := h^{++\alpha}_{a=1}$ and  $h^{+++} := h^{+++}_{a=1}$.
The corresponding gauge transformations are given by
\begin{equation} \label{eq: GF}
\delta_\lambda h^{++\alpha} = \mathcal{D}^{++} \lambda^\alpha
    +
    2 \theta^{+\alpha} \lambda^+\,,
    \qquad
    \delta_\lambda h^{+++} = \mathcal{D}^{++}\lambda^+\,,
\end{equation}
which are just the $a=1$ subset of \eqref{eq: GFa}. Using  this gauge freedom, one can choose Wess-Zumino type gauge:
\begin{equation}\label{WZ}
    \begin{split}
        h^{++\alpha}
        =&\;
        (\bar{\theta}^+)^2 \psi^{\alpha} - 4i \theta^{+\beta}\bar{\theta}^{+\dot{\beta}} \Psi_{\beta\dot{\beta}}^\alpha
        +
        8i
        (\theta^+)^2 \bar{\theta}^{+\dot{\beta}} C_{\dot{\beta}}^{\alpha i} u^-_i
        \\&
        +
        (\bar{\theta}^+)^2 \theta^{+\alpha} F^i u^-_i
        +
        (\bar{\theta}^+)^2 \theta^{+\beta} \mathcal{F}_{(\beta}^{\alpha)i} u^-_i
        +
        (\theta^+)^4 D^{\alpha(ij)}u^-_i u^-_j,
        \\
        h^{+++} =&\;
        (\theta^+)^2 \bar{\theta}^{+\dot{\beta}} \bar{\rho}_{\dot{\beta}}
        -4
        (\bar{\theta}^+)^2 \theta^{+\beta} \kappa_\beta
        +
        (\theta^+)^4 C^i u^-_i.
    \end{split}
\end{equation}
The residual gauge freedom is spanned by the parameters:
\begin{equation}
    \begin{split}
        \lambda^\alpha =&\; \epsilon^\alpha
    +
    2 \theta^{+\alpha} c^i u^-_i,
    \\
    \lambda^+ =&\;
    c^i u^+_i
    +
    2i
    \bar{\theta}^{+\dot{\beta}} \bar{\eta}_{\dot{\beta}}
    +
    4i
    \theta^{+\rho}\bar{\theta}^{+\dot{\rho}} \partial_{\rho\dot{\rho}} c^i u^-_i.
    \end{split}
\end{equation}
The residual gauge transformations act on the fields in WZ gauge as\footnote{We use the definition $\partial_{\alpha\dot{\alpha}}
= \frac{1}{2} \sigma^m_{\alpha\dot{\alpha}} \partial_m$, so $\partial_{\alpha\dot{\beta}}\partial^{\dot{\beta}\beta} = \frac{1}{4}\delta_\alpha^\beta \Box$.}:
\begin{equation}
    \delta_\lambda C_{\beta\dot{\beta}}^{ i} = \partial_{\beta\dot{\beta}} c^i,
    \qquad
    \delta_\lambda C^i  = 2 \Box c^i,
    \qquad
    \delta_\lambda \Psi_{\beta\dot{\beta}}^\alpha = \partial_{\beta\dot{\beta}} \epsilon^\alpha
    +
    \delta_\rho^\beta \bar{\eta}_{\dot{\rho}},
    \qquad
    \delta_\lambda \kappa_\beta = \partial_\beta^{\dot{\beta}} \bar{\eta}_{\dot{\beta}}.
\end{equation}
These transformation laws mean that $C^i_{\beta\dot{\beta}}$ is the doublet of spin 1 gauge fields, $\Psi^\alpha_{\beta\dot{\beta}}$ is the conformal gravitino field,
see, e.g., \cite{Fradkin:1985am}. Like in $\mathcal{N}=2$ supergravity \cite{Ivanov:2024gjo}, other fields with non-trivial gauge transformations can be redefined with the help
of the gauge fields so as to be inert under the gauge group. So they become the auxiliary fields. In particular, the bosonic field  $C^i$  can be redefined as
\begin{equation}
    C^i = 4 \partial^{\beta\dot{\beta}} C_{\beta\dot{\beta}}^i
    +
    G^i\,,
\end{equation}
where $G^i$ is a field invariant under gauge transformations. For the conformal gravitino, using the parameter $\bar{\eta}_{\dot{\alpha}}$ (which can be interpreted as a parameter of local conformal supersymmetry),
one can impose the symmetric gauge on the spin $\frac{3}{2}$ field,  $\Psi_{\beta\rho\dot{\rho}} = \Psi_{(\rho\beta)\dot{\rho}}$.
In this gauge, $\bar{\eta}_{\dot{\rho}} = - \frac{1}{2}\partial_{\rho\dot{\rho}} \epsilon^{\rho} $, and we are left with the gauge transformations:
\begin{equation}
    \delta\Psi_{(\alpha\beta)
        \dot{\alpha}}
    =
    \partial_{(\alpha\dot{\alpha}} \epsilon_{\beta)},
    \qquad
    \delta \kappa_\beta = \partial_\beta^{\dot{\beta}} \bar{\eta}_{\dot{\beta}}
    =
    \frac{1}{2}
    \partial_\beta^{\dot{\beta}} \partial_{\dot{\beta}}^\gamma \epsilon_\gamma
    =
    -
    \frac{1}{8} \Box \epsilon_\beta.
\end{equation}
The proper redefinition leads to the fermionic field $\chi_\beta$ that is inert under gauge transformations:
\begin{equation}
    \delta\left(\partial^{\alpha\dot{\alpha}} \Psi_{(\alpha\beta)\dot{\alpha}} \right)
    =
    \frac{3}{4} \Box \epsilon_\beta
    \qquad
    \Rightarrow
    \qquad
    \kappa_\beta = - \frac{1}{6} \left(\partial^{\alpha\dot{\alpha}} \Psi_{(\alpha\beta)\dot{\alpha}} \right)
    +
    \chi_\beta.
\end{equation}

As the result, we derive the off-shell contents of $\mathcal{N}=2$ conformal gravitino
supermultiplet as\footnote{In theories with higher derivatives some auxiliary fields also become dynamical
(see, e.g. \cite{Ivanov:2024gjo}) and there is no natural way to distinguish between physical and auxiliary fields.}:
\begin{equation}
    \begin{split}
    \text{Gauge sector:}& \qquad \Psi_{(\alpha\beta)\dot{\alpha}},
         C^i_{\alpha\dot{\alpha}};
        \\
        \text{Non-gauge sector:}& \qquad
         \psi_\alpha,
        \rho_\alpha, \chi_\alpha, G^i, F^i,
        \mathcal{F}_{}^{(\alpha\beta)i},
        D^{\alpha(ij)}.
    \end{split}
\end{equation}
Thus we finally have $\mathbf{32_B}+ \mathbf{32_F}$ off-shell degrees of freedom. We fix the canonical dimension
for the gravitino field as $[\Psi_{(\alpha\beta)\dot{\alpha}}] = \frac{3}{2}$,  then the vector field has the non-canonical dimension
$[C^i_{\alpha\dot{\alpha}}]=2$, which is typical for $R$-symmetry gauge field in conformal supergravity models. Respectively, $[\kappa] = -1$ and the prepotentials
have dimensions $[h^{++\alpha}] = \frac{1}{2}$, $[h^{+++}]=1$.

This dimensional analysis implies  that the free gauge-invariant component action of $\mathcal{N}=2$ gravitino multiplet contains the conformal gravitino
action involving three derivatives, together with the standard  Maxwell action for the doublet of  vector gauge fields.
The gauge-invariant action for  conformal gravitino can be  constructed in terms of the gauge-invariant strengths:
\begin{equation}\label{eq: spin 3/2 gi}
    \check{C}_{(\dot{\alpha}\dot{\beta}\dot{\gamma})}
    :=
    \partial_{(\dot{\alpha}}^{\alpha} \partial_{\dot{\beta}}^\beta \Psi_{(\alpha\beta)\dot{\gamma})},
    \qquad
    \hat{C}_{(\alpha\beta\gamma)}
    :=
    \partial_{(\alpha}^{\dot{\rho}} \Psi_{\beta\gamma)\dot{\rho}},
\end{equation}
\begin{equation}
    S_{conf}^{\frac{3}{2}} = \int d^4x \;\left( \hat{C}^{(\alpha\beta\gamma)} \check{\bar{C}}_{(\alpha\beta\gamma)}
    +
    \check{C}_{(\dot{\alpha}\dot{\beta}\dot{\gamma})} \hat{\bar{C}}^{(\dot{\alpha}\dot{\beta}\dot{\gamma})}   \right).
\end{equation}
Here we follow the same notations for the strengths as in \cite{Kuzenko:2019ill}.

\section{Invariant action}

To construct an invariant action, we need to pass to the superfields with the scalar transformation laws under supersymmetry (analogously to the case of integer $\mathcal{N}=2$ higher spins \cite{Buchbinder:2021ite}).
The analytic prepotential $h^{++\alpha}$ has the nontrivial supersymmetry transformation law \eqref{eq: susy transf}. The superfields transforming under supersymmetry as scalars  are defined by
\begin{equation}\label{eq: covar prepot}
    G^{++\alpha} = h^{++\alpha} + 2\theta^{-\alpha} h^{+++},
    \qquad
    G^{+++} = h^{+++}.
\end{equation}
The tilde-conjugated superfields are defined as $\bar{G}^{++\dot{\alpha}} = \widetilde{G^{++\alpha+}}$ and $\bar{G}^{+++} = \widetilde{G^{+++}}$.

The newly defined $G^{++}$-superfields possess the gauge transformation laws
\begin{equation}\label{eq: cov GF}
    \delta_\lambda G^{++\alpha} = \mathcal{D}^{++} \Lambda^\alpha,
    \qquad
    \delta_\lambda G^{+++}
    =
    \mathcal{D}^{++} \Lambda^+,
\end{equation}
with
\begin{equation}\label{eq: Lambda parameters}
    \Lambda^\alpha = \lambda^\alpha
    +
     2\theta^{-\alpha} \lambda^{+},
    \qquad
    \Lambda^+ = \lambda^+.
\end{equation}

The set of negatively charged gauge  prepotentials is introduced as solutions of the zero-curvature harmonic conditions:
\begin{equation}\label{eq: zero curvature eq}
    \begin{split}
    &\mathcal{D}^{++} G^{--\alpha} = \mathcal{D}^{--} G^{++\alpha},
    \\
    &\mathcal{D}^{++} G^{--+} = \mathcal{D}^{--} G^{+++},
    \\
    &\mathcal{D}^{++} G^{---} = G^{--+}.
    \end{split}
\end{equation}
The requirement of covariance of these equations under the gauge transformations \eqref{eq: cov GF} fixes the gauge transformation laws of the negatively charged potentials:
\begin{equation}
    \begin{split}
    &\delta_\lambda G^{--\alpha} = \mathcal{D}^{--} \Lambda^\alpha,
    \\
    &\delta_\lambda G^{--+} =
    \mathcal{D}^{--} \Lambda^+ - \Lambda^-,
    \\
    &\delta_\lambda G^{---} = \mathcal{D}^{--} \Lambda^-,
    \end{split}
\end{equation}
where the parameter $\Lambda^-$ is defined by the relation $\mathcal{D}^{++} \Lambda^- = \Lambda^+$.

Making use of the negatively charged potentials, we can construct two gauge-invariant $\mathcal{N}=2$ super strengths\footnote{We define  covariant derivatives as
$\mathcal{D}^+_{\hat{\alpha}} = \partial^+_{\hat{\alpha}},
\mathcal{D}^-_\alpha = - \partial^-_\alpha + 4i \bar{\theta}^{-\dot{\alpha}}\partial_{\alpha\dot{\alpha}},
\bar{\mathcal{D}}^-_{\dot{\alpha}} = - \partial^-_{\dot{\alpha}} - 4i \theta^{-\alpha}\partial_{\alpha\dot{\alpha}}$.}:
\begin{equation}\label{eq: SFS}
    \hat{\mathcal{W}}_\alpha = (\bar{\mathcal{D}}^+)^2 G^{--}_{\alpha},
    \qquad
    \check{\bar{\mathcal{W}}}_\alpha = (\bar{\mathcal{D}}^+)^2
     \left[
     \partial_{\alpha\dot{\alpha}} \bar{G}^{--\dot{\alpha}}
    +
    \frac{i}{2} \mathcal{D}^-_\alpha \bar{G}^{--+}
    -
    \frac{i}{2} \mathcal{D}^+_\alpha \bar{G}^{---}
    \right].
\end{equation}

The second superfield strength has an interesting geometric structure analogous to that of the linearized $\mathcal{N}=2$ super Weyl tensor \cite{Ivanov:2024gjo}.
Using this resemblance, we can transform this superstrength as follows.

We start with the composite object satisfying the ``half-analyticity'' condition:
\begin{equation}
    \bar{\mathcal{H}}^{++}_\alpha = \partial_{\alpha\dot{\alpha}}\bar{G}^{++\dot{\alpha}} + \frac{i}{2} \mathcal{D}^-_\alpha \bar{G}^{+++},
    \qquad
    \bar{\mathcal{D}}^+_{\dot{\alpha}}  \bar{\mathcal{H}}^{++}_\alpha = 0.
\end{equation}
Its gauge transformation is defined by:
\begin{equation}
    \delta_\lambda  \bar{\mathcal{H}}^{++}_\alpha = \mathcal{D}^{++} \bar{L}_\alpha,
    \qquad
    \bar{L}_\alpha = \partial_{\alpha\dot{\alpha}} \bar{\Lambda}^{\dot{\alpha}}
    +
    \frac{i}{2} \mathcal{D}^-_\alpha \bar{\Lambda};
    \qquad
    \bar{\mathcal{D}}^+_{\dot{\alpha}}  \bar{L}_\alpha = 0.
\end{equation}
Using the zero-curvature condition, one constructs the quantity $\bar{\mathcal{H}}^{--}_\alpha$
\begin{equation}
    \mathcal{D}^{++} \bar{\mathcal{H}}^{--}_\alpha
    =
    \mathcal{D}^{--} \bar{\mathcal{H}}^{++}_\alpha
    \quad
    \Rightarrow
    \quad
    \bar{\mathcal{H}}^{--}_\alpha = \partial_{\alpha\dot{\alpha}} \bar{G}^{--\dot{\alpha}}
    +
     \frac{i}{2} \mathcal{D}^-_\alpha \bar{G}^{--+}
    -
    \frac{i}{2} \mathcal{D}^+_\alpha \bar{G}^{---}
\end{equation}
with  the gauge transformation law
\begin{equation}
    \delta \bar{\mathcal{H}}^{--}_\alpha = \mathcal{D}^{--} \bar{L}_\alpha.
\end{equation}
Now, using $ \mathcal{H}^{--}_\alpha$, we can construct the gauge invariant
\begin{equation}
    \check{\bar{\mathcal{W}}}_\alpha
    =
    (\bar{\mathcal{D}}^+)^2 \bar{\mathcal{H}}^{--}_\alpha.
\end{equation}
Its gauge invariance is just the consequence of the ``half-analyticity'' of  $\bar{L}_\alpha$.

\medskip

The superstrengths $\hat{\mathcal{W}}^\alpha$ and $\check{\bar{\mathcal{W}}}_\alpha$  reveal some  useful properties:

\medskip

$\bullet$ \textit{\underline{Harmonic independence}}:
\begin{equation}\label{HarInd}
        \mathcal{D}^{\pm\pm}    \hat{\mathcal{W}}_\alpha = 0,
        \qquad
    \mathcal{D}^{\pm\pm}    \check{\bar{\mathcal{W}}}_\alpha = 0.
\end{equation}
 These equations can be easily checked, for example,
 \begin{equation}\label{HarInd 1}
    \mathcal{D}^{++} \check{\bar{\mathcal{W}}}_\alpha
    =
    \mathcal{D}^{++} (\bar{\mathcal{D}}^+)^2 \bar{\mathcal{H}}^{--}_\alpha
    =
     (\bar{\mathcal{D}}^+)^2
        \mathcal{D}^{--} \bar{\mathcal{H}}^{++}_\alpha
        =
        8i \theta^{-\rho} \partial_{\rho\dot{\rho}} \bar{\mathcal{D}}^{+\dot{\rho}} \bar{\mathcal{H}}^{++}_\alpha = 0,
 \end{equation}
 where we have used the ``half-analyticity'' property. From \eqref{HarInd 1} it also follows that $   \mathcal{D}^{--} \check{\bar{\mathcal{W}}}_\alpha = 0$.

 \medskip

 $\bullet$ \textit{\underline{Chirality}}:
 \begin{equation}
    \bar{\mathcal{D}}^{\pm}_{\dot{\alpha}}  \hat{\mathcal{W}}_\alpha = 0,
    \qquad
    \bar{\mathcal{\mathcal{D}}}^{\pm}_{\dot{\alpha}}    \check{\bar{\mathcal{W}}}_\alpha = 0.
 \end{equation}
 The conditions $\bar{\mathcal{D}}^+_{\dot{\alpha}}\hat{\mathcal{W}}_\alpha = 0$ and $\bar{\mathcal{D}}^+_{\dot{\alpha}}\check{\bar{\mathcal{W}}}_\alpha = 0$
 follow from the definitions  \eqref{eq: SFS}. The conditions  $\bar{\mathcal{D}}^-_{\dot{\alpha}}\hat{\mathcal{W}}_\alpha = 0$
 and $\bar{\mathcal{D}}^-_{\dot{\alpha}}\check{\bar{\mathcal{W}}}_\alpha = 0$ are consequences of the harmonic independence condition:
 \begin{equation}
    \mathcal{D}^{++} \left( \bar{\mathcal{D}}^-_{\dot{\alpha}}\hat{\mathcal{W}}_\alpha\right) = 0
    \qquad
    \Rightarrow
        \qquad
    \bar{\mathcal{D}}^-_{\dot{\alpha}}\hat{\mathcal{W}}_\alpha = 0.
 \end{equation}

 The component contents of $\mathcal{N}=2$ superfield strengths \eqref{eq: SFS} can be displayed by expressing the negatively charged prepotentials
 as solutions of the zero curvature conditions \eqref{eq: zero curvature eq} in W.Z. gauge.
 However, in order to determine, up to numerical  coefficients, the component structure in the gauge field sector it suffices to make use of
 the requirement of gauge invariance, chirality and harmonic independence.

 For the conformal gravitino one can construct two gauge invariants \eqref{eq: spin 3/2 gi} with the dimensions $[\hat{C}_{(\alpha\beta\gamma)}] = \frac{5}{2}$ and
 $[\check{C}_{(\dot{\alpha}\dot{\beta}\dot{\gamma})}] = \frac{7}{2}$. For the spin $1$ gauge field one can construct two gauge invariants with dimensions
 $[\mathcal{F}^i_{(\alpha\beta)}] = [\bar{\mathcal{F}}^i_{(\dot{\alpha}\dot{\beta})}]=2$, which extend the self-dual and  anti self-dual parts of the Maxwell tensor:
  \begin{equation}
    \mathcal{F}^i_{(\alpha\beta)} = \partial_{(\alpha}^{\dot{\beta}} C^i_{\beta)\dot{\beta}},
    \qquad
    \mathcal{F}^i_{(\dot{\alpha}\dot{\beta})}
    =
    \partial_{(\dot{\alpha}}^\beta C^i_{\beta\dot{\beta})}.
  \end{equation}

  Since $[\hat{\mathcal{W}}_\alpha] = \frac{3}{2}$, $[\check{\bar{\mathcal{W}}}_{\dot{\alpha}}] = \frac{5}{2}$, the gauge field sectors of these objects (up to numerical factors)
  are uniquely fixed as:
 \begin{subequations}
    \begin{equation}
        \hat{\mathcal{W}}_\alpha
        \sim
        \theta^{+(\beta}\theta^{-\gamma)} \hat{C}_{(\alpha\beta\gamma)}+
        (\theta^-)^2 \theta^{+\beta} \mathcal{F}_{(\alpha\beta)}^i u^+_i
        +
        (\theta^+)^2 \theta^{-\beta} \mathcal{F}_{(\alpha\beta)}^i u^-_i + \dots,
    \end{equation}
    \begin{equation}
        \check{\bar{\mathcal{W}}}_\alpha
        \sim
        \theta^{+\beta} \bar{\mathcal{F}}_{(\alpha\beta)}^i u^-_i
        -
        \theta^{-\beta} \bar{\mathcal{F}}_{(\alpha\beta)}^i u^+_i
        +
        \theta^{+(\beta}\theta^{-\gamma)} \check{\bar{C}}_{(\alpha\beta\gamma)} + \dots.
    \end{equation}
 \end{subequations}
So $\hat{\mathcal{W}}_\alpha$ and $\check{\bar{\mathcal{W}}}_\alpha$ are  $\mathcal{N}=2$ supersymmetric extensions of
the gauge invariant tensors $\hat{C}_{(\alpha\beta\gamma)}$ and  $\check{\bar{C}}_{(\alpha\beta\gamma)} $.

\medskip

Using chirality of $\hat{\mathcal{W}}^\alpha$ and $\check{\bar{\mathcal{W}}}_\alpha$, we now can construct $\mathcal{N}=2$ supersymmetry invariant
action as  an integral over $\mathcal{N}=2$ chiral superspace:
\begin{equation}\label{eq: 3/2 action}
    S = \int d^4x d^4 \theta \; \hat{\mathcal{W}}^\alpha \check{\bar{\mathcal{W}}}_\alpha + c.c.
\end{equation}
In the gauge field sector, this action yields (up to numerical coefficients):
\begin{equation}
    S = \int d^4x\, \left( \hat{C}^{(\alpha\beta\gamma)} \check{\bar{C}}_{(\alpha\beta\gamma)}
    +
    \mathcal{F}^{(\alpha\beta)}_i \bar{\mathcal{F}}^i_{(\alpha\beta)}  \right) + c.c.
\end{equation}

\medskip

Finally, we demonstrate that the  action  \eqref{eq: 3/2 action} can be rewritten as an integral over the full $\mathcal{N}=2$ harmonic superspace:
\begin{equation}\label{eq: harmonic action}
    S = \int d^4x d^8\theta du\;
    \left[ G^{++\alpha} \bar{\mathcal{H}}^{--}_\alpha
    -
    \bar{G}^{++}_{\dot{\alpha}}
    \mathcal{H}^{--\dot{\alpha}}
    \right].
\end{equation}
Using the definitions \eqref{eq: SFS} and the properties of $\mathcal{N}=2$ superstrengths, it can be checked that
\begin{equation}
    \begin{split}
    \int d^4x d^8\theta du\;  G^{++\alpha} \bar{\mathcal{H}}^{--}_\alpha
    =&
    \,
    \frac{1}{16} \int d^4x d^4 \theta du \, (\bar{\mathcal{D}}^+)^2 (\bar{\mathcal{D}}^-)^2
    \left[ G^{++\alpha} \bar{\mathcal{H}}^{--}_\alpha \right]
    \\=&\,
    \frac{1}{16} \int d^4x d^4 \theta du \,  (\bar{\mathcal{D}}^-)^2
G^{++\alpha} (\bar{\mathcal{D}}^+)^2 \bar{\mathcal{H}}^{--}_\alpha
\\=&\,
\frac{1}{16} \int d^4x d^4 \theta du \,  (\bar{\mathcal{D}}^-)^2
G^{++\alpha} \check{\bar{\mathcal{W}}}_\alpha.
    \end{split}
\end{equation}
Further, making use of the relation
\begin{equation}
    (\bar{\mathcal{D}}^-)^2
    G^{++\alpha}
    =
    \mathcal{D}^{--} \left[     (\bar{\mathcal{D}}^+\bar{\mathcal{D}}^-)    G^{++\alpha} \right]
    -
    (\bar{\mathcal{D}}^+\bar{\mathcal{D}}^-)    \mathcal{D}^{++} G^{--\alpha}\,,
\end{equation}
we obtain
\begin{equation}
        \int d^4x d^8\theta du\;  G^{++\alpha} \bar{\mathcal{H}}^{--}_\alpha
        =
        \frac{1}{16} \int d^4x d^4\theta du \,
        (\bar{\mathcal{D}}^+)^2
        G^{--\alpha} \check{\bar{\mathcal{W}}}_\alpha
        =
        \frac{1}{16}
        \int d^4x d^4\theta \,
        \hat{\mathcal{W}}^\alpha \check{\bar{\mathcal{W}}}_\alpha.
\end{equation}
Though the harmonic superspace action \eqref{eq: harmonic action} lost manifest gauge invariance (it is rather of Chern-Simons type), it has the structure
typical for the analogous harmonic superspace actions of other gauge supermultiplets. Using the properties $\int  G^{++\alpha} \bar{\mathcal{H}}^{--}_\alpha
= \int G^{--\alpha} \bar{\mathcal{H}}^{++}_\alpha $ and $ \int \bar{G}^{++}_{\dot{\alpha}}
\mathcal{H}^{--\dot{\alpha}} = \int \bar{G}^{--}_{\dot{\alpha}}
\mathcal{H}^{++\dot{\alpha}}  $,
one can also obtain some equivalent harmonic representations  of the action  \eqref{eq: 3/2 action}.

\section{Mezincescu-type prepotential}\label{eq: sec 6}

In the case of $\mathcal{N}=2$ supergravity \cite{Zupnik:1998td, Kuzenko:1999pi, Butter:2010sc} and $\mathcal{N}=2$ integer higher spin theories \cite{Buchbinder:2022vra},
there exists an alternative description of the underlying degrees of freedom in terms of the Mezincescu-type gauge prepotential. In particular, this description
is  convenient for constructing various $\mathcal{N}=2$ cubic interactions of higher spins \cite{Zaigraev:2024ryg}. It is useful to consider a similar representation
for the $\mathcal{N}=2$ conformal gravitino supermultiplet.

The invariant $\mathcal{N}=2$ superfields \eqref{eq: covar prepot} satisfy the constraints:
\begin{equation}
    \begin{cases}
        \mathcal{D}^+_{\hat{\alpha}} G^{+++} = 0,
        \\
        \mathcal{D}^+_\beta G^{++\alpha} =  2 \delta^\alpha_\beta G^{+++},
        \\
            \bar{\mathcal{D}}^+_{\dot{\beta}} G^{++\alpha} = 0.
    \end{cases}
\end{equation}
These constraints can be resolved in terms of the single unconstrained Mezincescu-type prepotential $\Upsilon^-(\zeta, u)$
\begin{equation}
    \begin{cases}
     G^{+++} = (\mathcal{D}^+)^4 \Upsilon^-,
     \\
     G^{++\alpha} = - \frac{1}{4} \mathcal{D}^{+\alpha} (\bar{\mathcal{D}}^+)^2 \Upsilon^-,
     \end{cases}
\end{equation}
which is defined up to the pre-gauge freedom,
\begin{equation}\label{eq: pre gauge}
    \delta_B \Upsilon^- = (\mathcal{D}^+)^2 B^{(-3)}  + \bar{\mathcal{D}}^+_{\dot{\alpha}} B^{(-2)\dot{\alpha}}.
\end{equation}

The original gauge freedom \eqref{eq: cov GF} corresponds to the following transformations of the Mezincescu prepotential:
\begin{equation}\label{eq: Y harm transf}
    \delta_\lambda \Upsilon^- = \mathcal{D}^{++} K^{(-3)},
\end{equation}
so that the transformation $\Lambda$-parameters defined in \eqref{eq: Lambda parameters} are expressed as:
\begin{equation}
        \Lambda^+ = (\mathcal{D}^+)^4   K^{(-3)},
    \qquad
    \Lambda^{\alpha} = - \frac{1}{4} \mathcal{D}^{+\alpha} (\mathcal{D}^+)^2 K^{(-3)}.
\end{equation}
It follows from this discussion, that in the expansion of $\Upsilon^-$ over the analytic superfields there remain few terms,
\begin{equation}
    \Upsilon^- (\theta^-, \zeta) = (\theta^-)^4 h^{+++}(\zeta)
    +
    (\bar{\theta}^-)^2 \theta^-_\alpha h^{++\alpha}(\zeta)
    +
    \text{pre-gauge d.o.f.},
\end{equation}
which are recognized as the analytic prepotentials \eqref{eq: GF}. All other analytic superfields in this expansion
can be gauged away using the pre-gauge freedom \eqref{eq: pre gauge}.

Using the gauge transformation \eqref{eq: Y harm transf}, one can impose the gauge:
\begin{equation}
    \Upsilon^- (z,u)= \Upsilon^i(z) u^-_i,
\end{equation}
in which the remaining degrees of freedom are described by  the harmonic-independent superfield $\Upsilon^i(z)$,
with the residual gauge transformations
\begin{equation}\label{eq: harm ind GF}
    \delta_{\zeta, \omega} \Upsilon^i  = \mathcal{D}_{\alpha j} \zeta^{\alpha(ij)}
    +
    \bar{\mathcal{D}}_{\dot{\alpha} (j}\bar{\mathcal{D}}^{\dot{\alpha}}_{k)} \omega^{(ijk)}.
\end{equation}
In this way we have recovered the gauge superfield $\Upsilon_i$, firstly introduced in \cite{Hutchings:2023iza} (based on the earlier results of ref. \cite{Kuzenko:2021pqm}).
This superfield was introduced there as a gauge prepotential associated with $\mathcal{N}=2$ supercurrent obtained through $\mathcal{N}=2$ reduction
of  $\mathcal{N}=3$ conformal supercurrent \cite{Howe:1981qj}.  As a specific consequence of the supersymmetric reduction, this $\mathcal{N}=2$ supercurrent necessarily
couples to the $\mathcal{N}=2$ superconformal gravitino multiplet. So $\Upsilon_i$ can be identified with $\mathcal{N}=2$ conformal gravitino prepotential.
In this gauge, the action \eqref{eq: 3/2 action} reduces to the action of ref. \cite{Hutchings:2023iza}.

Analogously to the model of $\mathcal{N}=2$ vector multiplet \cite{Mezincescu}, the gauge freedom \eqref{eq: harm ind GF} of
the Mezincescu prepotential in the present case has no geometric interpretation (see also discussion in \cite{18}).
This is  in contrast with the gauge freedom realized on the analytic prepotentials  \eqref{eq: GF}.

\medskip
In terms of the Mezincescu prepotential $\Upsilon^-$ the cubic vertex \eqref{eq: cubic vertex} can be rewritten as the following integral over
the full $\mathcal{N}=2$ harmonic superspace:
\begin{equation}
    S_{int} = - \kappa \int d^4x d^8\theta du \; \left[ \Upsilon^{-a} \mathfrak{J}^+_a + \; \widetilde{} \;\;\; \text{conj}. \right],
\end{equation}
where $\mathfrak{J}^+_a$ is the $\mathcal{N}=2$ master supercurrent defined in \eqref{eq: principle sc}. This supercurrent satisfies the conservation
conditions  \eqref{eq: psc cons equation} which agree with the gauge transformations \eqref{eq: pre gauge} and \eqref{eq: Y harm transf} \cite{Zaigraev:2024ryg}.

\section{Harmonic geometry and supertwistors}

As was mentioned in section \ref{sec: intro}, the analytic prepotentials of $\mathcal{N}=2$ integer higher spins have a natural geometric interpretation
as vielbeins of the covariant derivative $\mathfrak{D}^{++}$ defined in \eqref{eq: Ein}, \eqref{eq: Conf} and generalizing $\mathcal{D}^{++}$ to the $\mathcal{N}=2$
supergravity case \cite{Buchbinder:2021ite, Buchbinder:2024pjm, Zaigraev:2024xve}.  No such an interpretation for $\mathcal{N}=2$ gravitino  prepotentials
$h^{++\alpha}, h^{+++}$ can be directly adduced, because otherwise it would be possible to construct a minimal interaction of the $\mathcal{N}=2$ gravitino
multiplet with the hypermultiplet, which is impossible.  Nonetheless, such an interpretation can be achieved by adding to the ${\cal N}=2$ harmonic analytic coordinates
some new auxiliary coordinates $\Psi^\alpha$ and $\omega^+$. Then, assuming analyticity and independence of vielbeins of these auxiliary coordinates, we introduce:
\begin{equation}\label{eq: gravitino}
    \mathfrak{D}^{++}_{gravitino} =
\mathcal{D}^{++} + h^{++\alpha} (\zeta) \frac{\partial}{\partial \Psi^\alpha}
+
h^{+++}(\zeta) \frac{\partial}{\partial \omega^+}
-
2 \theta^{+\alpha} \omega^+ \frac{\partial}{\partial \Psi^\alpha} + \;\;\widetilde{}\;\; \text{conj.}
\end{equation}
So $h^{++\alpha}$ is a veilbein corresponding to $\Psi^\alpha$, while $h^{+++}$ is a veilbein associated with $\omega^+$ (it is worth noting that one can double the number
of the auxiliary coordinated as $\Psi^{\alpha a}, \omega^{+a}$ and consider a doublet of $\mathcal{N}=2$ gravitino supermultiplet).

Gauge transformations can be generated using an analytic differential operator  associated with the auxiliary coordinate translations $\Psi^\alpha \to \Psi^\alpha + \lambda^\alpha$, $\omega^+ \to \omega^+ +\lambda^+$:
\begin{equation}
    \hat{\Lambda} = \lambda^\alpha \frac{\partial}{\partial \Psi^\alpha}
    +
    \lambda^+ \frac{\partial}{\partial \omega^+}
    + \;\;\widetilde{}\;\; \text{conj.}
\end{equation}
Then, in full analogy with $\mathcal{N}=2$ higher spins \cite{Buchbinder:2022kzl, Buchbinder:2022vra},
\begin{equation}
\delta_\lambda \mathfrak{D}^{++}_{gravitino} =  [   \mathcal{D}^{++}, \hat{\Lambda}]
    =
    \mathcal{D}^{++}  \lambda^\alpha \frac{\partial}{\partial \Psi^\alpha}
    +
    \mathcal{D}^{++}  \lambda^+ \frac{\partial}{\partial \omega^+}
    +
    2 \theta^{+\alpha} \lambda^+ \frac{\partial}{\partial \Psi^\alpha}.
\end{equation}
In such a way we reproduce the gauge freedom of the analytic prepotentials \eqref{eq: GF}.

The transformations of rigid $\mathcal{N}=2$ supersymmetry \eqref{eq: susy transf} can be reproduced by assuming the following transformation laws for the auxiliary coordinates:
\begin{equation}
    \delta_\epsilon \Psi^\alpha = - 2 \epsilon^{-\alpha} \omega^+,
    \qquad
    \delta_\epsilon \omega^+ = 0
\end{equation}
and requiring the invariance of the harmonic derivative:
\begin{equation}
    \delta_\epsilon \mathfrak{D}^{++}_{gravitino}  = 0.
\end{equation}

The possibility of such a representation for the new prepotentials hints at the possible harmonic origin of $\mathcal{N}=3$ and $\mathcal{N}=4$ supergravity prepotentials.
Indeed, the reduction of  $\mathcal{N}=3,4$ supergravity prepotentials to $\mathcal{N}=2$ should naturally lead to $\mathcal{N}=2$ supergravity and $\mathcal{N}=2$ gravitino prepotentials,
which correspond to the analytic vielbeins, as in \eqref{eq: Ein}, \eqref{eq: Conf} and in \eqref{eq: gravitino}.  So if we assume that the Grassmann coordinate $\Psi^\alpha$
can be interpreted as an additional odd coordinate in some $\mathcal{N}=3,4$ extended superspace and $\omega^+$ as $u^{+i}$ projection of some auxiliary bosonic coordinate $\omega^i,\, \omega^+ = \omega^iu^+_i$,
the $\mathcal{N}=2$ gravitino prepotentials are just the corresponding analytic vielbeins.  This reasoning leads to the conjecture that  $\mathcal{N}=3,4$
supergravity-covariant harmonic derivatives can be given the structure such that after reduction to $\mathcal{N}=2$ they reproduce the corresponding analytic prepotentials.
Perhaps, an adequate framework for such an interpretation is provided by the bi-harmonic formulation of ${\cal N}=4$ SYM theory (or a vector-hypermultiplet complex from the ${\cal N}=2$ supersymmetry point of view)
\cite{BuIvaIva2} in which both the ${\cal N}=2$ gauge superfield strength ${\cal W}$ appearing in \eqref{DefJalpha} and in subsequent formulas, as well as
the hypermultiplet superfield $q^{+a}$, are accommodated by a single bi-harmonic superfield strength.

\medskip
Alternatively, the additional coordinates can be interpreted as a sort of \textit{supertwistors}. If we consider the doublet $\mathcal{N}=2$ gravitino, then we need the doubled auxiliary coordinates $(\Psi^{\alpha a},
\omega^{+a})$. So we deal with a  doublet of spinors ($\Psi^{\alpha a}$)encompassing eight Grassmann coordinates, and four Lorentz scalars ($\omega^{ia}$). Surprisingly, these coordinates match the ``twisted'' $\mathcal{N}=2$ supertwistor, firstly introduced in
\cite{Ferber:1977qx}\footnote{Its components have the opposite Grassmann parities compared to the standard $\mathcal{N}=2$ supertwistor which consists of two even spinorial doublets and four odd scalar components. Possible
applications of the twisted supertwistor were discussed in \cite{Lukier}.}. Superconformal group $SU(2,2|2)$ have a  natural simple linear realization on supertwistors.  It would be very interesting to work out such an
analogy in more detail and to study the relevant superconformal properties of the $\mathcal{N}=2$ gravitino prepotentials introduced above (with respect to both ${\cal N}=2$ superconformal symmetry and its properly realized
${\cal N}=4$ extension $PSU(2,2|4)$).

\section{Conclusions}

In this paper, we have constructed, at the linearized level, the $\mathcal{N}=2$ off-shell multiplet containing  conformal gravitino field. It is described in terms of unconstrained analytic prepotentials $h^{++\alpha}$,
$h^{+++}$ and their tilde-conjugates. The component content of the multiplet was found in Wess-Zumino type gauge.  These $\mathcal{N}=2$ gravitino superfield prepotentials were deduced by the Noether method, through coupling
to the conserved $\mathcal{N}=2$ supercurrents associated with ``hidden'' symmetries of the free $\mathcal{N}=2$ vector-hypermultiplet system (linearized ${\cal N}=4$ Maxwell multiplet in the ${\cal N}=2$ superfield
formulation). These prepotentials and supercurrents determine the consistent cubic $(\mathbf{\frac{3}{2}}, \mathbf{1}, \mathbf{\frac{1}{2}})$ vertex in the form $prepotential \times supercurrent$.  Using the harmonic zero
curvature equations, we have determined two gauge-invariant chiral $\mathcal{N}=2$ superfield strengths $\hat{\mathcal{W}}_\alpha$ and $\check{\bar{\mathcal{W}}}_\alpha$ and, with their help,  constructed the invariant
action yielding the action of conformal gravitino after passing to components. We have presented the full harmonic superspace version of this action principle.

We have also found an alternative superfield representation of the involved degrees of freedom
in terms of a Mesincescu-type prepotential $\Upsilon^-$. In the harmonic-independent gauge it reduces to the prepotential $\Upsilon^{i}$,
recently introduced in \cite{Hutchings:2023iza}. To conclude, we have shown that the analytic $\mathcal{N}=2$ gravitino prepotentials
can be interpreted as additional  vielbeins of the covariant harmonic derivative $\mathfrak{D}^{++}$, provided that some auxiliary coordinates $\Psi^\alpha$
and $\omega^+$ were added. In such an extended frame, the gauge and $\mathcal{N}=2$ supersymmetric transformations of the extra prepotentials can be
derived quite similarly to the case of integer $\mathcal{N}=2$ higher spins.
We also discussed the closely related interpretations of the auxiliary coordinates as coordinates of some analytic $\mathcal{N}=3,4$ superspaces and/or  as Ferber's supertwistor.

\medskip

The most crucial open problem seems to be the construction of $\mathcal{N}=2$ Einstein gravitino multiplet and its action in the harmonic approach. We expect that, like in the case of $\mathcal{N}=2$ Einstein supergravity
\cite{Galperin:1987ek}, these can be deduced through couplings to the appropriate conformal compensators. In  the case at hand the natural candidates for compensators are $\mathcal{N}=2$ vector multiplet and hypermultiplet,
as is dictated by the vertex \eqref{eq: cubic vertex}. Alternative versions of $\mathcal{N}=2$ matter compensators could also be acceptable, analogously to the case $\mathcal{N}=2$ Einstein supergravity \cite{18,
Galperin:1987ek, Galperin:1986fg}. We hope to address these issues soon.

We expect that the approach presented here can be generalized to arbitrary $\mathcal{N}=2$ half-integer conformal spins.
Another interesting problem is to investigate the superconformal
properties of the resulting multiplet using the realization of superconformal group in the harmonic superspace \cite{18, Buchbinder:2024pjm, Galperin:1985zv}
or on supertwistors \cite{Ferber:1977qx}. It would also be interesting to explore models with an extended supersymmetry. For instance, one could study $\mathcal{N}=4$ conformal supergravity
represented in terms of $\mathcal{N}=2$ Weyl multiplet and $\mathcal{N}=2$ gravitino multiplet and examine how hidden $\mathcal{N}=4$ supersymmetry is realized in this model.

\acknowledgments


We are grateful to Yu. Zinoviev for pointing out the importance of studying higher-spin $\mathcal{N}=2$ supermultiplets coupled to $\mathcal{N}=2$ vector-hypermultiplet system.
N.Z. thanks Slava Ivanovskiy for the discussion
of   hidden (super)symmetries.
 Work of N.Z. was partially supported by
the Foundation for the Advancement of Theoretical Physics and
Mathematics ``BASIS''.



\end{document}